# Física, investigación científica y sociedad en la Argentina de 1930-1940

**Alejandro Gangui** | Instituto de Astronomía y Física del Espacio, CONICET - Universidad de Buenos Aires
gangui@iafe.uba.ar
https://orcid.org/0000-0002-4864-5348

**Eduardo L. Ortiz** | Imperial College, London
e.ortiz@imperial.ac.uk
https://orcid.org/0000-0001-8236-8770

**RESUMEN**
En este trabajo continuamos nuestro estudio sobre las investigaciones científicas llevadas a cabo en el Instituto de Física de la Universidad Nacional de La Plata durante la primera mitad del siglo XX, y sobre el importante papel que tuvo en ellas el físico Ramón G. Loyarte. A partir de sus trabajos, solo o en colaboración con miembros del Instituto, hacia fines de la década de 1920 Loyarte propuso la posible existencia de un nuevo cuanto de energía rotacional en el átomo de mercurio. Esta idea levantó críticas y generó una dura polémica científica que llegó a los medios académicos nacionales e internacionales. Analizamos aquí esta polémica mediante el estudio de los comentarios que aparecieron publicados en revistas internacionales de reseña. También estudiamos las actividades de Loyarte en la política partidaria y en la política de la ciencia en la Argentina de las décadas de 1930 y 1940, donde fue el responsable de interesantes iniciativas políticas y científicas.

**Palabras clave** Instituto de Física de La Plata - Ramón G. Loyarte - polémica Loyarte-Loedel Palumbo - revistas internacionales de reseña científica - política argentina en la primera mitad del s. XX.

**ABSTRACT**
*In this paper we continue our study of the scientific research carried out at the Institute of Physics of the National University of La Plata during the first half of the 20th century, and of the important role played by the physicist Ramón G. Loyarte. Based on his studies, alone or in collaboration with members of the Institute, towards the end of the 1920s Loyarte proposed the possible existence of a new quantum of rotational energy in the mercury atom. This idea raised criticism and generated a fierce scientific controversy that reached the national and international academic media. We analyse this controversy by studying the comments that appeared in international review journals. We also study Loyarte's activities in politics and science policy in Argentina in the 1930s and 1940s, where he was responsible for interesting political and scientific initiatives.*

# Introducción

En una primera parte de nuestro proyecto nos hemos concentrado en la figura del físico argentino Ramón G. Loyarte (1888-1944), quien fuera director del Instituto de Física de la Universidad Nacional de La Plata (UNLP) y, con el tiempo, llegara a ser presidente de dicha Universidad y miembro de la Academia de Ciencias de Buenos Aires (Gangui y Ortiz, 2022.1). Hemos detallado sus investigaciones y el rol que desempeñó en el avance de las ciencias durante la década de 1920. También presentamos sus trabajos científicos en áreas de la física cuántica experimental y su novedosa propuesta de la rotación cuantificada de los átomos (Loyarte, 1927.1; 1927.2).

En esta parte de nuestro trabajo hacemos foco en la *polémica* que surgió cuando una de sus líneas de investigación, quizás la más original, ligada con la introducción de un nuevo número cuántico, que pudo haber tenido una particular resonancia internacional, fue fuertemente rebatida. Su principal crítico fue otro miembro del Instituto de Física: Enrique Loedel Palumbo (1901-1962), un exalumno doctoral del profesor alemán Richard Gans (1880-1954) de cuya actuación en el campo de la investigación nos hemos ocupado más ampliamente en (Gangui y Ortiz, 2020). Esa crítica de las ideas de Loyarte, que como informamos en el presente trabajo no careció totalmente de antecedentes en Europa, terminó siendo localmente aceptada. El debate científico suscitado por esos trabajos de Loyarte y de Loedel Palumbo, localmente en la Argentina de esos años, ha dejado un registro limitado debido principalmente al desarrollo precario de los medios locales de difusión científica.

A partir de esa polémica, que coincide con el lanzamiento de su carrera política, la producción científica original de Loyarte disminuyó considerablemente. Sin embargo, no se agotó: la revista alemana *Physikalische Zeitschrift* y otras revistas destacadas continuaron publicando resultados de sus estudios, o de estudios suyos en colaboración con colegas del Instituto.

Los constantes intercambios de Loyarte con la vida social y política de su tiempo lo condujeron a ser elegido miembro del Parlamento Nacional en 1932; desde esa posición promovió interesantes iniciativas, algunas relacionadas con el avance de la ciencia en la Argentina que discutimos someramente. Consideramos también en este trabajo su actuación a partir de junio de 1943, después de que las fuerzas armadas quebraron la vida constitucional argentina por segunda vez en un lapso de menos de 15 años. Loyarte, identificado con ese movimiento, fue designado Interventor del Consejo Nacional de Educación, el organismo superior de supervisión de la enseñanza universal, laica y gratuita en las escuelas del Estado. La tarea asignada a Loyarte fue la de intentar, desde su alto cargo, desarmar el aparataje legal que, por más de medio siglo, había resguardado la libertad de culto dentro de la escuela argentina. Con su salud seriamente debilitada, Loyarte falleció sin haber podido cumplir con la difícil tarea que se le había confiado.

# 1. Una élite científica en conflicto: el Instituto de Física hacia fines de los años 1920

Como vimos en (Gangui y Ortiz, 2022.1), varios trabajos de Loyarte sugerían la existencia de un nuevo *cuanto* de energía rotacional, propuesta teórica que no escapó a la atención de los investigadores extranjeros, muchos de ellos consumados especialistas en espectroscopía y, en general, en física cuántica experimental. En esencia, Loyarte ofrecía evidencia experimental de que algunos potenciales de excitación del vapor de mercurio no correspondían a líneas espectrales conocidas. Y proponía que estas líneas se podían deducir por la adición o substracción de 1,4 volts (o de un múltiplo de esa cantidad) a potenciales ya conocidos.

Vimos también que la visita de Paul Langevin (1872-1946) a la Argentina en 1928 había sido prudentemente capitalizada por Loyarte, quien en sus publicaciones invocaba el nombre del prestigioso miembro del *Collège de France* como prueba de que sus investigaciones iban en la buena dirección.

## 1.1 Algunos comentarios de la hipótesis de Loyarte desde fuera de la Argentina

Si las suposiciones de Loyarte hubieran sido correctas, el nuevo número cuántico que él creía haber descubierto habría catapultado su nombre, sin duda, al escenario más selecto de la física cuántica de su época. Difícilmente Langevin, de quien hablamos en la primera parte de nuestro estudio (Gangui y Ortiz, 2022.1), podía haber dejado de notar esa posibilidad cuando escuchó noticias acerca de la dirección que habían tomado las investigaciones de Loyarte y de sus colaboradores en La Plata en 1928 y, gentilmente, hizo cálculos sobre el talio para ayudar a Loyarte en sus investigaciones (Loyarte, 1928, p. 228). Por otra parte, Langevin tampoco podía ignorar los trabajos fundamentales publicados unos años antes y, aproximadamente, en la misma dirección. Por ejemplo, los de George Uhlenbeck (1900-1988) y Samuel Goudsmit (1902-1978) sobre la rotación cuántica del electrón, que aparecieron más de dos años antes en revistas prestigiosas y ampliamente leídas, como *Die Naturwissenschaften* (Uhlenbeck y Goudsmit, 1925), o *Nature* (Uhlenbeck y Goudsmit, 1926).

Como sabemos, en esos años la Academia Nacional de Ciencias de Buenos Aires carecía aún de un órgano de difusión propio: por esa razón *Anales* de la Sociedad Científica Argentina (SCA), que anotaremos como ASCA, actuaba como el órgano de difusión de sus actividades. A través de una inteligente política de canje, promovida activamente por el matemático Claro Dassen (1873-1941), *Anales* habían alcanzado una circulación relativamente amplia en Europa. Consecuentemente, esa revista se había convertido en una plataforma adecuada para la difusión nacional e internacional de las ideas de los científicos locales: Loyarte fue uno de ellos. A la vez, esa política había ayudado a enriquecer la Biblioteca de periódicos de la SCA.

Hacia 1930, en momentos críticos para la historia de la Argentina, a los que nos referiremos más adelante, y pocos años después del comienzo de la publicación de la serie de notas sobre la rotación cuantificada, de Loyarte y de sus colaboradores, comenzaron a surgir indicadores de que sus ideas sobre un nuevo número cuántico no eran unánimemente aceptadas. Nos referiremos específicamente a una de ellas elegida, principalmente, por haberse originado dentro del círculo mismo de los investigadores del Instituto de Física de La Plata.

Antes de considerar esa divergencia recordemos que hacia 1931 los trabajos de Loyarte sobre la rotación cuantificada inicialmente encontraron un eco positivo. Por ejemplo, en Jean Genard,[1] uno de los principales redactores de *Ciel et Terre*, el Boletín de la Sociedad Belga de Astronomía. Después de leer noticias sobre los trabajos de Loyarte en ASCA (que, como acabamos de ver, circulaba en Europa), Genard escribió: '*L'existence d'une rotation quantifiée de l'électron constitue la révélation d'une fait jusqu'à présent inconnu, ainsi que l'existence de raies spectrales correspondantes*' (Genard, 1931.1). Sin embargo, en el número inmediatamente siguiente de esa misma revista, en una más meditada y técnica segunda nota, Genard señaló que Loyarte estaba equivocado al afirmar que con anterioridad a su trabajo los físicos no habían considerado la rotación de los electrones alrededor de ejes polares. Señaló también que con los dos parámetros cuánticos asociados con un nivel energético electrónico no es posible explicar la estructura fina de ciertas rayas espectrales, y que esta había sido la causa de que se introdujera una nueva subdivisión de niveles, lo que a su vez había requerido un nuevo número cuántico. A continuación, se refirió a los trabajos de los físicos holandeses Uhlenbeck y Goudsmit, publicados en 1925-26, que 'partiendo de hechos experimentales, [esto es, de] la estructura fina de las rayas espectrales', habían considerado ya la posibilidad de una rotación propia de los electrones alrededor de un eje que pasa por su centro de gravedad (Genard, 1931.2, p. 301). Genard destacó que no se trataba de una mera hipótesis ya que, dentro de los dos o tres años de haber sido formulada, la idea de un electrón rotante había sido adoptada, internacionalmente, por los especialistas en espectroscopía. Finalmente afirmó que: 'desde ya sería totalmente imposible omitirlas si se desea interpretar las líneas espectrales' (Genard, 1931.2, p. 302).

---

[1] Genard, un investigador activo, pertenecía al Departamento de Astrofísica de la Universidad de Liège.

Inmediatamente a continuación de esta segunda, y más sobria, evaluación del trabajo de Loyarte, apareció una nota breve del físico y erudito belga Eugène-Auguste Lagrange[2] (Lagrange, 1931) que, luego de agradecer la aclaración de Genard, lamentó que en su nota de ASCA Loyarte: 'ne citait pas même leurs noms [de Uhlenbeck y Goudsmit] et l'auteur soulignait même sa priorité'.

**1.2 La situación en la Argentina: la crítica de Loedel Palumbo**

Con anterioridad a ese breve desacuerdo, documentado en Bélgica en 1931 y quizás no el único en Europa, y de las notas publicadas en revistas de reseña de los Estados Unidos e Inglaterra que hemos discutido en (Gangui y Ortiz, 2022.1), hubo un comentario crítico local de los trabajos que Loyarte leyó ante la Academia y luego publicó en ASCA, en la revista *Contribución al estudio de las ciencias físicas y matemáticas* (que llamaremos *Contribución* en lo que sigue) y en *Physikalische Zeitschrift*. El autor de esa crítica fue Enrique Loedel Palumbo quien, agudamente, penetró en la estructura misma de los razonamientos de Loyarte.

En (Gangui y Ortiz, 2020) nos hemos referido ya a ese joven investigador, formado en el Instituto de Física de la UNLP, en la que en esos años se desempeñaba como profesor adjunto. Loedel Palumbo fue el último de los discípulos doctorales de Gans y en ocasión de la visita de Einstein a la Argentina en 1925, aún no graduado, tuvo un papel singular en las conversaciones que ese sabio sostuvo con un grupo elegido de físicos locales entre los que se contaba Loyarte (Ortiz, 1995; Gangui y Ortiz, 2008). Como también hemos visto en (Gangui y Ortiz, 2022.1), Loyarte lo eligió como su colaborador en la redacción de una celebrada obra de física elemental para la escuela secundaria (Loyarte y Loedel Palumbo, 1928; 1932), sin duda interesante y moderna. Hemos indicado también que durante su Presidencia de la UNLP Loyarte impulsó un plan de becas externas de las que Loedel Palumbo fue uno de los primeros beneficiarios (en los años 1928-1929).[3] Su futuro parecía promisorio.

En enero de 1930 Loedel Palumbo envió a la Secretaría de la Universidad, para su posible publicación en *Contribución,* un trabajo en el que criticaba la validez de los resultados de las investigaciones de Loyarte. Ese trabajo apareció publicado en esa revista un año más tarde (Loedel Palumbo, 1931). Esencialmente, Loedel Palumbo cuestionaba la realidad física, la unicidad y el significado de 'mínimo' del número propuesto por Loyarte. Además, presentaba conclusiones que diferían substancialmente de las que Loyarte había formulado en sus publicaciones especializadas, argentinas y extranjeras; entre éstas últimas, principalmente en *Physikalische Zeitschrift*. Su crítica fue puntual y se orientó a detallar inconsistencias estructurales, tanto en esos trabajos de Loyarte, como en los de Loyarte con algunos de sus colaboradores que, como Loedel Palumbo, eran todos miembros del Instituto de Física de La Plata.

---

[2] Eugène Lagrange (1855-1936), muy conocido por los lectores de esa revista, firmó su nota con sus iniciales: [E. L.]. Lagrange era un antiguo miembro del Consejo de Redacción de *Ciel et Terre*; en esos años, además, era presidente de la *Sociedad Belga de Astronomía.* Sus conocimientos enciclopédicos incluían el dominio de varios idiomas, uno de ellos el español.
[3] Ver Loedel Palumbo (1940; 1959).

Hemos señalado, brevemente, el desacuerdo suscitado en Bélgica en 1931 en conexión con los trabajos que Loyarte leyó en la Academia y que luego publicó en ASCA: no es imposible que a través de sus contactos con físicos alemanes Loedel Palumbo haya sido testigo de comentarios adversos a la línea de investigación que Loyarte desarrollaba en La Plata, lo que justificaría la confianza con la que abordó su detallado análisis de los diferentes resultados propuestos por Loyarte y sus colaboradores. Contrariamente a la segunda nota de Genard en *Ciel et Terre*, que como vimos apareció en 1931, Loedel Palumbo dejó totalmente de lado cuestiones relativas a las prioridades y centró su análisis en las premisas básicas sobre las que se apoyaban los trabajos de Loyarte (Loyarte, 1927.1; 1927.2). Esencialmente, mostró que las coincidencias entre los resultados experimentales y los deducidos sumando o restando un múltiplo de 1,4 volts no se debían a una causa atribuible a la física.

Su análisis de las tablas de resultados experimentales ofrecidas por Loyarte y sus colaboradores mostró que el número propuesto, los 1,4 volts, no era el único valor que permitía obtener las 'coincidencias' señaladas por Loyarte: prácticamente cualquier otro número podría reemplazar a ese valor. La existencia de esos múltiples divisores se debería, esencialmente, a que esa muestra de datos numéricos era adecuadamente extensa. La sugerencia, incluida en su trabajo, según la cual Loyarte había sido selectivo en el uso de los datos numéricos derivados de sus experimentos es contenciosa e innecesaria, dado el peso de sus argumentos.

Ampliando el escenario de la discusión desde La Plata a Alemania (posiblemente debido a la demora en la aparición de *Contribución*) Loedel Palumbo envió una versión en alemán de, esencialmente, el mismo trabajo a *Physikalische Zeitschrift*. Esa revista acusó recibo de su envío el 18 de febrero de 1930, y publicó la nota hacia fines de ese mismo año (Loedel Palumbo, 1930), con anterioridad a su aparición en *Contribución*, en Argentina.

Sin duda sorprende la ductilidad de los equipos de selección de contribuciones en ambas revistas: con igual indiferencia publicaron resultados que afirmaban o que negaban la misma premisa. Además, esa premisa era nada menos que la existencia de un nuevo número cuántico. Quizás sorprende aún más el caso de *Physikalische Zeitschrift*, una revista científicamente cotizada y considerada, con razón, como una publicación seria y responsable en la selección de las contribuciones que admitía en sus páginas. Indudablemente fue indulgente con su colega en La Plata.

En el caso de La Plata asombra la ausencia de una discusión amplia dentro de las estructuras del Instituto de Física: una discusión técnica dentro de su Seminario hubiera permitido, sin duda, aclarar sus diferencias a su debido tiempo. La designación especial de una *Sala de Seminarios* había sido ya prevista por Emil Bose en el rediseño original del edificio del Instituto de Física (Ortiz, 2021) y, muy tempranamente, comenzó a utilizarse con ese propósito específico. Gans, el segundo director del Instituto de Física, continuó prestando atención a ese primer escalón de discusión, de alto valor educativo. Sin duda, en un centro de investigaciones de las dimensiones del de La Plata, el debate científico interno debía ser una parte esencial en el proceso de entrenamiento de su personal científico. Sin embargo, como ha señalado Pyenson (Pyenson, 1985, pp. 237-38), ese primer filtro de los resultados científicos alcanzados en el Instituto fue abandonado durante la dirección de Loyarte.

Lamentablemente el examen crítico de Loedel Palumbo, que dejaba pocas dudas, no cerró el debate: el camino erróneo de investigación en el que el Instituto de Física se había embarcado, con tan sugestiva unanimidad, no se detuvo en 1930. Aunque las críticas de Loedel Palumbo contribuyeron a desacelerar esa línea de trabajo, de hecho, en forma muy pronunciada, la publicación de varios trabajos posteriores, que no tienen en cuenta sus reservas, sugiere que la crítica de Loedel Palumbo fue aceptada con cierta dificultad.

El equipo y las técnicas experimentales utilizadas por Loyarte y sus colaboradores, y la crítica de Loedel Palumbo, han sido consideradas en detalle en (von Reichenbach y Andrini, 2015), particularmente en pp. 173-79, donde se discute también una posible dimensión ética en el debate suscitado por Loedel Palumbo con Loyarte, en p. 183.

**1.3 Ecos de la polémica sobre la rotación cuántica de Loyarte en la Argentina**

Con excepción de *Contribución*, revista local donde había aparecido la crítica de Loedel Palumbo a los trabajos de Loyarte, en esos años no existía una revista argentina específicamente dedicada a la investigación en física, menos aún al análisis de trabajos originales en esa disciplina. En cambio, se publicaba regularmente una revista que incluía una sección dedicada a la crítica de estudios recientes relativos a las ciencias exactas: el *Boletín del Seminario Matemático Argentino*, dirigido por Julio Rey Pastor (1888-1962). Este *Boletín* contaba con el auspicio de la Facultad de Ciencias de la Universidad de Buenos Aires y fue un antecesor directo de la *Revista de la Unión Matemática Argentina*, que funcionó también como órgano de difusión de estudios de los físicos locales (Ortiz, 2016).

En la sección de bibliografía del *Boletín* apareció una referencia breve al trabajo de Loedel Palumbo. En pocas líneas y con precisión, se describieron las tesis de los principales trabajos de Loyarte y sus colaboradores, dando preferencia a las ideas de Loedel Palumbo. Con respecto al nuevo número cuántico, el autor de la referencia (cuya calidad y precisión sugiere que puede haber sido el mismo Rey Pastor) destacó que Loedel Palumbo: 'Demuestra que tal número carece de realidad física y, aplicando el cálculo de probabilidades, deduce que cualquier otro número [en lugar de 1,4 volts] puede justificarse análogamente' (*Boletín del Seminario Matemático Argentino*, 1932, p. 74).

**1.4 Algunos comentarios de la polémica entre Loyarte y Loedel Palumbo fuera de la Argentina**

El impacto de las observaciones críticas de Loedel Palumbo acerca de las ideas de Loyarte, relativas a una posible rotación cuantificada del átomo de mercurio, no se limitaron a la Argentina: fueron también recogidas por varias revistas internacionales de referencia que hemos citado en (Gangui y Ortiz, 2022.1).

El destacado especialista en espectroscopía, W. Stiles, que como hemos indicado antes había reseñado los trabajos de varios físicos platenses para la revista *Science Abstracts – Physics* se ocupó, en (Stiles, 1931.1), de la nota crítica que Loedel Palumbo había publicado en *Physikalische Zeitschrift* en 1930 (Loedel Palumbo, 1930). Indicó que Loedel Palumbo criticaba la hipótesis de Loyarte acerca de la rotación cuantificada de los átomos de mercurio, talio y potasio, y mostraba 'que la concordancia entre los potenciales críticos calculados y observados y longitudes de onda ópticas es enteramente fortuita'. Agregó que en ese trabajo 'se muestra que se puede obtener un acuerdo igualmente bueno si, en el caso del mercurio, se elige como 'cuanto rotacional', cualquier otro valor, totalmente arbitrario y 'diferente de los 1.4 volts de Loyarte, …, digamos 1 volt'. Finalmente señaló que 'en ninguno de los casos ligados con el mercurio, talio o potasio se puede atribuir significación física alguna a las numerosas relaciones [numéricas] observadas por Loyarte'.

A continuación de su reseña de la nota de Loedel Palumbo, Stiles (Stiles, 1931.2) se ocupó de la nota en la que Loyarte se defiende de las críticas (Loyarte, 1930) publicada también en *Physikalische Zeitschrift*. Stiles indicó que su autor reseñaba la evidencia experimental acerca de la existencia de una rotación cuantificada de 1,4 volts para el átomo de mercurio, concluyendo que 'todos los potenciales críticos de origen desconocido se pueden generar sumando o restando 1.4 volts, o un múltiplo simple de esa cantidad, a partir de los potenciales críticos calculados de niveles ópticos'. La estructura del argumento de Stiles es análoga al juicio crítico, muy preciso, ofrecido por el *Boletín del Seminario Matemático Argentino*.

En 1932 *Science Abstracts - Physics* también se hizo eco de la nota de Loedel Palumbo, en su versión de 1931 en *Contribución* (Loedel Palumbo, 1931). El autor de esta reseña, H. F. Gillbe (Gillbe, 1932.1), indicó que el trabajo de Loedel Palumbo demostraba que la hipótesis de Loyarte acerca de las rotaciones cuantificadas del átomo, y especialmente de los átomos de mercurio, talio y potasio, 'carece de fundamento, ya que la concordancia [numérica] en que se basa es meramente fortuita'. Señaló, además, que 'el llamado *potencial de adición-substracción* no tiene significado físico' y que 'las mediciones del potencial crítico del mercurio deben considerarse como carentes de valor'.

Gillbe reseñó, también en *Science Abstracts – Physics*, la respuesta de Loyarte al artículo de Loedel Palumbo (Gillbe, 1932.2), nuevamente desde la versión publicada en *Contribución* (Loyarte, 1931.2). Indicó que Loyarte argumentaba que la existencia del potencial de 1,4 volts se podía deducir, no sólo de sus propias observaciones, sino también de las tablas de observaciones realizadas por *otros* autores, lo que no contestaba la crítica formulada por Loedel Palumbo. Señaló también que Loyarte indicaba que los potenciales que él había calculado servían, también, para explicar resultados experimentales de H. Messenger (Messenger, 1926).

Por otra parte, W. R. Angus[4] pertenecía entonces a la misma escuela de espectroscopía que Richard A. Morton (pionero de la espectroscopía en investigaciones bioquímicas y que ya había reseñado trabajos más convencionales de Loyarte) y, hacia 1930, reseñaba también para *British Chemical Abstracts*: en esa revista ya había considerado otros trabajos de Loyarte y de sus colaboradores. En 1931 se ocupó de la polémica de Loedel Palumbo y Loyarte desde la plataforma de *Physikalische Zeitschrift*. Primeramente, consideró la nota de Loedel Palumbo (Loedel Palumbo, 1930) en (Angus, 1931.1). Escuetamente señaló que, según ese autor, 'el "potencial de adición", postulado por Loyarte no tiene significación física alguna y que [Loedel Palumbo] presenta una tabla para el espectro del mercurio que muestra que, para este "potencial de adición", se pueden utilizar otros valores *diferentes de 1.4 volts* [las itálicas son nuestras], obteniéndose con ellos los mismos resultados que deduce Loyarte; y que en el caso de cualquier otro espectro que contenga un número [suficientemente] grande de líneas se puede hacer este cambio'.[5] Su evaluación del argumento de Loedel Palumbo coincide, nuevamente, con el juicio crítico, muy preciso, del *Boletín del Seminario Matemático Argentino*.

En la misma revista, y a continuación de la referencia anterior, Angus (Angus, 1931.2) reseñó la respuesta que Loyarte había dado en *Physikalische Zeitschrift* (Loyarte, 1930) a las críticas de Loedel Palumbo. Señaló que Loyarte afirmaba nuevamente que el origen de ciertas líneas ópticas del espectro del mercurio se podría explicar sumando o restando 1,4 volts (o múltiplos pequeños de ese número) del valor de la energía potencial de una línea de origen conocido. Agregó que en ese trabajo Loyarte analizaba también el *significado físico* de ese procedimiento y que, con un número de tablas de datos

---

[4] W. Rogy Angus era profesor en el University College, Gales del Norte, Bangor.

[5] "The 'addition potential' postulated by Loyarte has no physical significance. A table is given for the mercury spectrum showing that values *other than 1.4 volts* [las itálicas son nuestras] for this 'addition potential' can be used to give the same results as Loyarte obtained. This can be done for any spectrum which contains a large number of lines".

experimentales, trataba de mostrar el buen acuerdo entre los valores calculados y los observados.

A pesar de esta difícil polémica, que alcanzó nivel internacional, Loyarte no abandonó totalmente el tema de investigación que la había desatado: más adelante publicó otros trabajos que, en alguna forma, intentaban demostrar que con ciertas modificaciones se podrían refutar, o al menos soslayar, los argumentos de Loedel Palumbo y retener lo esencial de sus ideas. Interesa señalar que este *affaire* tampoco cerró su acceso a *Physikalische Zeitschrift*: en 1933, bajo su sólo nombre, publicó en esa revista alemana una nota sobre temas de espectroscopía (Loyarte, 1933); el contenido de esa nota no estaba totalmente desligado de sus antiguos estudios sobre el equívoco potencial de 1,4 volts.

## 1.5 El potencial crítico de Loyarte y la investigación internacional contemporánea

En las líneas que siguen trataremos, muy esquemáticamente, de posicionar los conceptos científicos de Loyarte y de sus colaboradores dentro del mundo de ideas de su época. Hemos visto que algunos temas de investigación considerados por Loyarte eran compartidos por otros autores a nivel internacional, aunque no necesariamente participando con las ideas del físico platense, o compartiendo sus conclusiones. Algunas de esas consideraciones pueden ayudarnos a comprender más objetivamente *la polémica de los 1,4 volts*.

En esos años el estudio de las descargas eléctricas en gases continuaba siendo un tema de gran interés, tanto fundamental como tecnológico. Joseph John Thomson (1856-1940) lo consideró globalmente en una obra clásica publicada en Inglaterra en el cambio de siglo. Décadas más tarde, en 1928-33, en un período crítico para los estudios de Loyarte, aquella obra fue actualizada por su autor con la colaboración con su hijo, George P. Thomson (1892-1975) (Thomson y Thomson, 1933). La nueva y extensa versión se publicó en dos volúmenes: en el primero los autores consideraron la ionización por calor y luz, y en el segundo se ocuparon de la ionización por colisión y por descargas en gases. Dada la calidad de estos investigadores (padre e hijo recibieron el Premio Nobel de Física) y a pesar de la amplitud que adquirió ese tema, el 'Thomson and Thomson' era una obra de referencia obligada en la década de la polémica de Loyarte y, también, luego de ella.

En esa obra se hace referencia a tópicos considerados por Loyarte y sus colaboradores. Sin embargo, no se cita específicamente ninguno de los trabajos de los investigadores platenses sobre esos temas. En cambio, se hace referencia a estudios paralelos realizados por otros autores: Thomson y Thomson señalaron que: 'En el caso del mercurio Franck y Einsporn, y también Pavlov y Sueva,[6] han observado [niveles espectroscópicos] críticos que no se pueden explicar con ningún nivel espectroscópico conocido. En algunos casos pueden deberse a moléculas inestables, en otros a transiciones desde un estado metaestable más bajo'. Más adelante agregan que: 'En algunos casos pueden ser debidos a errores experimentales. Discrepancias similares ocurren con uno o dos de los otros elementos'. (Thomson y Thomson, 1933, II, p. 79).

Esas últimas consideraciones destacan las dificultades teóricas y experimentales asociadas con los estudios en esas áreas de la física y pueden ayudarnos a comprender el debate platense dentro de una perspectiva más amplia. Desde luego muestran que Loyarte y sus colaboradores no estaban solos en el área de sus búsquedas científicas: el estudio de los potenciales críticos, central dentro del grupo de trabajos de Loyarte, continuó siendo un objeto de interés en la literatura científica de los primeros años de la década de 1930.

Aunque Loyarte, y también su frecuente colaborador Williams, eran investigadores experimentados, hay referencias a que algunas de sus exploraciones en este tema específico no siempre fueron compartidas por otros físicos contemporáneos. Algunos investigadores, por ejemplo, Filippov y Prokofjew (Filippov y Prokofjew, 1933, p. 657), señalaron en *Zeitschrift für Physik* que habían encontrado dificultades en detectar algunas líneas del espectro del Talio que Loyarte y Williams anunciaban haber observado. En esa misma revista y en esos mismos años otros investigadores citaron trabajos experimentales de Loyarte como una referencia a sus propios estudios; por ejemplo (Beutler y Demeter, 1934).

---

[6] Se refiere a (Pavlov y Sueva, 1929).

En 1932, es decir, contemporáneamente con la polémica sobre el potencial de 1,4 volts de Loyarte, el investigador estadounidense E. N. Shawhan[7] consideró el debate de la radiación correspondiente a potenciales críticos por debajo de los 4,66 volt (Shawhan, 1932) y atribuyó a Charles W. Jarvis (Jarvis, 1926), y más tarde a Pavlov y Sueva (Pavlov y Sueva, 1929), el origen de esos estudios. Agregó que no le había sido posible encontrar potenciales por debajo de 4,9 volts. El año siguiente el mismo Shawhan, ahora en colaboración con C. W. Jarvis, publicó una nota breve en *Physical Review* (Jarvis y Shawhan, 1933) en la que se discutió la posibilidad de descubrir radiaciones en el vapor de mercurio a potenciales críticos bajos. En esa nota se hizo referencia al hecho de que, en el caso del mercurio, diversos autores habían observado potenciales críticos bajos (por debajo de los 4,7 volts), citando los trabajos clásicos de 'Jarvis, Nielson, Pavlov y Sueva, y otros'. En ese mismo trabajo se ofrece una indicación rara y elocuente de la percepción contemporánea acerca de la validez de las ideas de Loyarte:

> 'R. G. Loyarte [PZ **30**. 923, 1929][8] attempts an explanation (*not well received*),[9] on the base of the rotation of the mercury atom as a free axial rotator, whose energy values change by multiples of 1.39 volts'.

En una nota posterior de Loyarte y Heiberg de Bose (Loyarte y Heiberg de Bose, 1935, pp. 24-25), los autores intentaron ofrecer una explicación a la falta de corroboración de algunos de los resultados anunciados por Loyarte y sus colaboradores, circunstancia a la que hemos hecho referencia específica antes. Su explicación tiene puntos en común con una de las posibilidades abiertas por Thomson y Thomson: las dificultades experimentales asociadas con esos estudios. Dicen Loyarte y Heiberg de Bose en (Loyarte y Heiberg de Bose, 1935) que han 'efectuado un gran número de medidas por el método fotoeléctrico. Hemos comprobado que la probabilidad de las transiciones atómicas depende de la intensidad de la corriente de electrones, esto es, de la densidad de las cargas espaciales. Potenciales críticos que aparecen nítidamente para corrientes débiles desaparecen o se hacen casi imperceptibles para corrientes intensas y recíprocamente.[10] *Esta es, quizás, la razón por la cual diversos observadores no encuentran los mismos potenciales. Este hecho señala la necesidad de hacer observaciones con diferentes cargas espaciales y a presiones diversas*' [las itálicas son nuestras]. En su producción posterior, al referirse a la posible significación especial del potencial de los 1,4 volts, Loyarte adoptó una posición de mucha mayor prudencia.

Otros trabajos de Loyarte, o de Loyarte con colaboradores, tuvieron mejor aceptación y siguieron siendo citados hasta una fecha relativamente reciente. Por ejemplo, la nota publicada por Loyarte y Williams en *Physikalische Zeitschirift* en 1929, (Loyarte y Williams, 1929), fue citada hasta por lo menos 1964, cuando Choong Shin-Piaw y Wang Loong-Seng (Shin-Piaw y Loong-Seng, 1964) recordaron que Loyarte y Williams fueron los primeros investigadores que estudiaron el espectro de líneas de absorción del vapor de la plata.

---

[7] Shawhan se desempeñaba entonces en el Mendenhall Laboratory of Physics, The Ohio State University.
[8] Se refiere a (Loyarte, 1929).
[9] Las itálicas son nuestras.
[10] Una década antes, en 1922, una estudiante de G.P. Thomson había hecho observaciones que apuntaban en esa misma dirección (Thomson y Thomson, 1933).

Williams, por su parte, continuó publicando resultados de sus estudios en revistas de prestigio utilizando la experiencia que había adquirido en el campo de la espectroscopía; en ocasiones se ocupó también de trabajos iniciados por Gans en La Plata. En las décadas de 1930 y 1940 este destacado investigador mantuvo correspondencia científica con figuras principales de la espectroscopía contemporánea.

### 1.6 Secuelas del conflicto: los trabajos posteriores de Loedel Palumbo y su futuro institucional incierto

La polémica generada dentro del Instituto de Física de la UNLP alrededor de un posible nuevo número cuántico y la dramática intervención de Loedel Palumbo no parece haber ayudado al avance de la promisoria carrera académica de este joven investigador. Es posible conjeturar que, en el pasado, las relaciones entre Loyarte y Loedel Palumbo fueron cordiales: sabemos que colaboraron en la redacción de una conocida obra de texto (Loyarte y Loedel Palumbo, 1928; 1932) y también que, como resultado de una provechosa iniciativa de Loyarte como presidente de la UNLP, Loedel Palumbo recibió una de las nuevas y muy codiciadas becas externas de esa universidad. Esa beca le permitió trasladarse a Alemania y realizar estudios avanzados de física y de filosofía; difícilmente la hubiera recibido sin el asentimiento de Loyarte, en su doble carácter de presidente de la Universidad y director del Instituto de Física. Sin embargo, a partir de 1930 las relaciones entre ambos no parecen haber sido las mismas que en el pasado.

El 14 de diciembre de 1931, luego de la polémica de 1929-30 acerca del supuesto nuevo número cuántico de Loyarte, Loedel Palumbo sometió a *Contribución* un nuevo e interesante trabajo sobre la estructura fina del espectro del hidrógeno.[11] En esos años difíciles *Contribución* se publicaba con un retraso considerable: el estudio de Loedel Palumbo recién apareció publicado en 1935 (Loedel Palumbo, 1935). Años más tarde, analizando la evolución de la física en la Argentina, entre 1924 y 1974, Westerkamp indicó que con ese trabajo 'el autor empieza a mostrarse como un agudo físico teórico' (Westerkamp, 1975, p. 5). Sorprende que este trabajo no haya tenido una versión correlativa en *Physikalische Zeitschrift* o en alguna de las grandes revistas europeas de circulación internacional, ocasionalmente utilizadas por investigadores del Instituto de Física.

Esta polémica de 1929-30 tuvo consecuencias importantes para el joven y prometedor físico teórico. En publicaciones posteriores a 1930 Loedel Palumbo volvió a ocuparse de temas de la teoría de la relatividad, donde ha dejado contribuciones de singular interés: en particular, los hoy llamados *diagramas de Loedel Palumbo* (Loedel Palumbo, 1948). Sin embargo, aun esa interesante contribución circuló, únicamente, a través de publicaciones locales: solamente su re-descubrimiento independiente, por parte de Henri Amar (Amar, 1955) y la difusión que tuvo a través de *Physical Review*, sumada a la integridad de Amar, que reconoció sin reservas la prioridad del físico platense (Amar, 1957), determinó que esos diagramas hoy lleven su nombre.

Pareciera, además, que ese joven investigador comprendió que su futuro profesional en La Plata tenía mejores perspectivas fuera del Instituto de Física: conservó su cátedra en la Facultad de Ciencias Fisicomatemáticas de la UNLP, pero estrechó su antiguo contacto con la Facultad de Humanidades de esa universidad, donde contribuyó decididamente a la formación de varias generaciones de futuros docentes de física, utilizando métodos originales y modernos. Desde el Colegio secundario y el Liceo de Señoritas anexos a la Universidad contribuyó, muy directamente, a enseñar física a alumnos que, muchos años después, lo recordarían con respeto.

---

[11] El Dr. Jorge Miraglia, investigador del Conicet en la Argentina y experto en física atómica y molecular, nos ha señalado que considera que este trabajo de Loedel Palumbo "es un artículo creativo y moderno para su época" (comunicación personal, 2020).

Loedel Palumbo publicó también estudios de valor sobre temas de pedagogía de las ciencias, en particular de la física. En una de sus obras más apreciadas consideró con agudeza los problemas que presenta la transmisión de los resultados de la física moderna a los estudiantes del nivel secundario. En otra parte hemos señalado la influencia considerable que tuvieron sus libros de texto en el proceso de modernización de la enseñanza de la Física a nivel secundario en la Argentina (Gangui y Ortiz, 2022.2).

## 2. La actividad de Loyarte en la política partidaria y en la política de la ciencia en la Argentina de 1930 - 1940

### 2.1 El clima de violencia política en la Argentina de principios de la década de 1930

En el giro de la década de 1920 a la de 1930 emergieron otros géneros de diferencias entre los científicos de la UNLP, diferente de las divergencias científicas surgidas en el campo específico de la investigación y a las que acabamos de referirnos. Quizás podría conjeturarse que esos dos tipos de diferencias no hayan estado totalmente desconectados.

En septiembre de 1930, por primera vez en la historia independiente de la Argentina, una sublevación militar, llamada desde entonces un *golpe de Estado militar*, derrocó al gobierno legalmente elegido. Ese golpe de Estado provocó una intensa radicalización en el país, particularmente manifiesta en el mundo intelectual. En el ambiente específico del Instituto de Física de la UNLP esa fractura colocó a algunos de sus principales actores a uno u otro lado del conflicto político. Loedel Palumbo, lo mismo que un buen número de otros físicos jóvenes de ese período: Rafael Grinfeld y Ernesto Sábato entre ellos, se opusieron con determinación al golpe de estado de 1930 y se acercaron a grupos políticos que trabajaban activamente por el retorno al régimen constitucional. A medida que avanzaba la década de 1930 la radicalización iniciada por ese golpe se agudizó a causa de la creciente complejidad de la situación política internacional, contribuyendo a profundizar las diferencias entre las ideas de aquel grupo de jóvenes y las de Loyarte.

Loyarte era un antiguo y activo miembro del tradicional Partido Conservador,[12] al que representó como diputado y al que continuó apoyando hasta el final de su vida. Desde fines de la década de 1920 Loyarte favorecía entusiastamente la idea de un golpe de Estado militar: ese golpe permitiría entregar el gobierno nacional a un grupo político selecto al que él pertenecía y que, muy difícilmente, lograría alcanzarlo a través de un proceso electoral democrático. Cuando el golpe de Estado ocurrió, el 6 de septiembre de 1930, el Partido Conservador lo apoyó decididamente; Loyarte hizo lo mismo.

En ese momento Loyarte había concluido ya su período como presidente de la UNLP y había retornado a sus tareas en el Instituto de Física. Su sucesor legal en la presidencia de esa universidad había sido el conocido escritor e historiador Dr. Ricardo Rojas (1882-1957), pero luego del golpe de Estado militar fue obligado a renunciar. El 25 de junio de 1931 el gobierno militar encargó la conducción de los destinos de la UNLP al Dr. Federico Walker, a quién dio el título ambiguo de *Interventor*. Con anterioridad este profesor de química había tratado de ocupar democráticamente la presidencia de la UNLP, pero había sido derrotado por Loyarte por un margen muy amplio.

---

[12] Ese partido tenía entonces dos ramas paralelas, Partido Conservador y Partido Provincial, fusionadas en el llamado Partido Demócrata Nacional a partir de 1931.

Aún un año después del golpe, hacia fines de 1931, Walker no había logrado modificar la situación dentro de la Universidad, que seguía siendo confusa y sumamente inestable. En septiembre de ese año hizo una crítica devastadora de la labor de los institutos y, en general, de la investigación en la UNLP; implícitamente, también de la gestión de Loyarte.

Sin dejar de atender a la polémica sobre la rotación cuantificada del átomo de mercurio, a la que nos referimos más atrás, Loyarte se vio forzado a abrir otro frente más para contestar los ataques de Walker. En su enérgica defensa reseñó las ideas que habían inspirado su actuación como presidente de la UNLP y envió copia de su carta al entonces ministro de Justicia e Instrucción Pública, Dr. Guillermo Rothe (1879-1959) (*El Argentino*, 1931.1). Loyarte se quejó de que 'el informe del Sr. Interventor se ha difundido por la prensa; [y, en consecuencia] constituye un documento oficial que pasará luego a los archivos', lo que comprometerá su 'buen nombre'. Enseguida rebatió las afirmaciones de Walker, que había expresado que las fundaciones institucionales de Loyarte fueron 'actos electoralistas', cuya consecuencia fue la creación de organismos 'raros e inútiles' que no podían 'aportar ningún beneficio a la enseñanza y a la investigación'. Con respecto a la creación de Departamentos dentro de la Universidad, una de sus innovaciones organizativas, Loyarte indicó que, efectivamente, el 15 de marzo de 1928 había enviado una nota a los Decanos sugiriendo la necesidad de modificar la estructura interna de las Facultades y la creación de Departamentos. Loyarte señaló también en su nota que creía que, al igual que en otras universidades europeas y americanas (entre las primeras citaba a la Universidad de Göttingen), es esencial 'que existan en la Universidad [argentina] posibilidades económicas de vida que respondan a necesidades científicas, o docentes'. Con esas iniciativas Loyarte trataba de reorientar la UNLP hacia la eventual creación de posiciones con dedicación exclusiva (o tiempo completo) insistiendo en la importancia educativa 'del ejemplo de la propia obra'; es decir, afirmando la figura del profesor que enseña pero que, además, investiga sobre lo que enseña. Loyarte terminó su carta a Walker con una lista de los principales trabajos realizados en el Museo, en el Observatorio y en otros institutos de la UNLP durante su mandato.

Refiriéndose a los resultados de las investigaciones realizadas en La Plata, Loyarte adoptó un punto de vista moderado, temperado quizás por sus pasadas experiencias en ese campo: 'No son, desde luego, grandes descubrimientos, a los cuales sólo se llega por una evolución lenta de la cultura. Estos trabajos, relativamente modestos, son un primer paso, el cual requiere, sin embargo, gran dominio de la ciencia a que se refieren y son prueba irrefutable de dedicación'.

A continuación, indicó que informaría 'sobre esa labor científica' a la Academia de Ciencias de Buenos Aires, de la que como sabemos era Miembro Titular y también secretario. Agregó que 'mi información será objetiva, limitada al aspecto científico y, en manera alguna, ha de referirse al informe del señor Interventor'. Efectivamente Loyarte produjo un informe sobre la labor realizada durante su presidencia de la UNLP, limitándose a los departamentos de Química y de Química Biológica y al Instituto de Anatomía de la Facultad de Medicina Veterinaria; luego lo comunicó a la Academia de Ciencias de Buenos Aires como había prometido.

Loyarte defendió también su pasada decisión de escuchar la voz y el voto de los representantes estudiantiles en las discusiones del Consejo Superior universitario, como lo establecía el reglamento universitario que él creyó que debía respetar. Señaló, además, que su relación con esos representantes había sido de mutuo respeto: 'mientras fui presidente me obedecieron y respetaron'.

Dos días después el Dr. Walker produjo una resolución por la cual 'se apercibe seriamente al señor profesor doctor Ramón G. Loyarte' por haber leído en la Academia una comunicación acerca de 'Institutos o Departamentos de esta Universidad que, dice, han sido suprimidos, lo que constituye una falta de disciplina' (*El Argentino,* 1931.2). También en 1931, Loyarte dejó constancia de sus esfuerzos publicando una recopilación de sus discursos, exposiciones e iniciativas durante su ejercicio de la presidencia de la Universidad de La Plata (Loyarte, 1931.1).

En una gestión paralela a sus ataques a Loyarte, Walker expulsó de la UNLP a unos 30 alumnos y a varios profesores, algunos de ellos conocidos nacional e internacionalmente.[13] La dimensión de las resistencias que Walker logró generar en sólo siete meses de gestión, dentro y fuera de la Universidad, es sorprendente: la situación de la UNLP aconsejó a las autoridades militares que, finalmente, debían deshacerse de él y buscar otros caminos.

Ante crecientes dificultades y reveses en diferentes frentes, lentamente, algunos miembros de ese gobierno militar comenzaron a comprender la complejidad de la función de gobierno. En el caso de la UNLP acudieron a no otro que Loyarte para que les ayudara a salir de la encrucijada terminando con los conflictos y pacificando la institución, sabiendo que se debía pagar un precio político. A mediados de enero de 1932[14] el gobierno militar designó a Loyarte Interventor de la Universidad que antes lo había elegido, democráticamente, como su presidente. Éste intentó conducir la UNLP, pero el clima de hostilidad que heredó de Walker había contribuido a generar una oposición intensa y relativamente unificada; tanto el Consejo Superior universitario como los profesores y los estudiantes se mostraron reacios a conciliar sin garantías. Luego de un período muy breve Loyarte comprendió que no le sería fácil seguir adelante y a mediados de mayo del mismo año presentó su renuncia.[15]

## 2.2 La actividad pública de Loyarte a principios de la década de 1930

Poco más tarde, repuesto el sistema constitucional, Loyarte volvió a la política y fue elegido Diputado ante el Congreso Nacional por el período 1932-34; representaba a la Provincia de Buenos Aires como parlamentario del Partido Conservador. Lo acompañaba otra figura de prestigio profesional, el cirujano Dr. José Arce (1881-1968),[16] que antes había sido Rector de la Universidad de Buenos Aires. Según (Béjar, 2005) Arce y Loyarte no pertenecían al sector partidario que tenía una inserción política sólida, o una influencia profunda en la base de ese movimiento: ambos eran, más bien, figuras conservadoras tradicionales que aportaban su prestigio personal o profesional a ese movimiento.

Sin embargo, en esos años estas dos figuras, como el resto de la población argentina, difícilmente pueden haber ignorado la existencia de una política de violencia e intimidación que ayudaba a sustentar a su partido en posiciones de poder. Ese clima no se limitaba a los políticos de base, o a personajes secundarios o poco conocidos dentro del mundo político argentino: también afectaba a dirigentes de envergadura nacional. El asesinato del Senador Nacional Enzo Bordabehere (1889-1935), opositor al gobierno y representante de un partido democrático, que ocurrió en 1935 dentro del recinto mismo de la Cámara de Senadores, es un ejemplo suficientemente elocuente de la existencia de un clima de violencia aplicado aun en los niveles políticos más elevados.

En sus intervenciones ante el Congreso Nacional Loyarte sostuvo posiciones constructivas, pero también apoyó otras que le acarrearon fuertes críticas. Entre las primeras citemos su defensa de los derechos de los trabajadores ferroviarios jubilados y del futuro de su Caja de Jubilaciones y Pensiones. También su apoyo a la propuesta de creación de una Dirección Nacional de Asistencia Social del Niño, basada en las ideas del prestigioso pediatra Dr. Gregorio Aráoz Alfaro (1870-1955). Finalmente, participó en las discusiones parlamentarias, y luego en una comisión específica, que condujo a la formulación de un sistema de protección de los derechos de autor. Loyarte percibía este último derecho como parte de la defensa del 'régimen de la propiedad privada' y como un 'reconocimiento del desarrollo y del florecer de la ciencia, de la técnica, de las letras y de las artes argentinas' (Diario de Sesiones, 1933). Con el decidido apoyo de Loyarte, el 28 de septiembre de 1933 se promulgó la Ley No. 11.723, o *Régimen legal de la propiedad intelectual*, que protege los derechos de autor (*Boletín Oficial*, 1933).

---

[13] Entre otros, Walker separó de la UNLP al destacado químico orgánico Dr. Enrique V. Zappi a quien, en 1928, el presidente Loyarte había designado jefe del Departamento de Bioquímica, el primero de una serie de nuevos Departamentos de la UNLP.
[14] El 15 de enero de 1932.
[15] El 15 de mayo de 1932.
[16] Diputado en el período 1934-36.

Entre sus posiciones antidemocráticas cabe mencionar su ambigüedad frente a los proyectos de supresión del sufragio universal obligatorio en la Argentina dando apoyo, hacia 1932-33, a iniciativas tendientes a la instauración de una forma selectiva y limitativa del sufragio; el llamado *voto calificado*. Éste último proponía la exclusión de personas, o de grupos específicos de personas, por razones de índole política, social o, incluso, en base a argumentos fundados en los principios de la llamada 'higiene social'.

## 2.3 Loyarte y el debate sobre la *nacionalización* de la dirección de institutos científicos en la Argentina

Recordemos que a mediados de la década de 1920 Loyarte ocupó la dirección del Instituto de Física de La Plata, sucediendo a dos directores extranjeros: Bose y Gans y, también, a un primer director-organizador uruguayo-argentino, el ingeniero Ricaldoni. Hacia principios de la década de 1930 es posible detectar el desarrollo de un complejo debate en el que se argumenta acerca de la necesidad de pasar la dirección de algunas de las grandes instituciones científicas de la Argentina a manos de investigadores locales; de ese debate (Ortiz, 2015) sólo nos interesa rescatar la parte que concierne a las ciencias exactas**.**

En el grupo de los países al sur del continente americano esa polémica no fue exclusiva de la Argentina. Un debate similar tuvo lugar, por ejemplo, en Chile donde el astrónomo alemán Friedrich Wilhelm Ristenpart (1868-1913), director de su principal Observatorio, enfatizó en diversas ocasiones la importancia que él mismo había atribuido a la formación de especialistas locales. Efectivamente, después de su trágico fallecimiento lo sucedió un astrónomo chileno que él mismo había entrenado.

En esos años, el Observatorio Astronómico Nacional, en Córdoba, posiblemente el más elaborado y antiguo establecimiento dedicado al cultivo de la ciencia pura en la Argentina, fue el foco principal de discusiones sobre estos temas. Desde su creación en 1871, cuando se contrató a su director-fundador, el Dr. Benjamín Gould (1824-1896), el Observatorio de Córdoba estuvo permanentemente bajo la dirección de astrónomos estadounidenses con la sola excepción de cortos períodos de espera entre dos directores sucesivos de esa misma nacionalidad. Si bien en sus primeros años este era un hecho auspicioso, ya que no había especialistas locales, luego de medio siglo de funcionamiento de ese instituto, algunos pensaban que era razonable esperar un cambio de escenario. En el primer cuarto del siglo XX uno de los más severos críticos de esa situación fue el geodesta y astrónomo Félix Aguilar (1884-1943).[17] Este destacado científico argentino criticaba al Observatorio, y también a otras instituciones, por su inhabilidad para formar un número adecuado de especialistas locales que pudieran luego insertarse en la institución y llegar a adquirir experiencia suficiente como para eventualmente orientarla. Es también cierto que esas inquietudes no siempre estuvieron acompañadas por un esfuerzo paralelo del Estado nacional por crear condiciones de vida y de trabajo adecuadas para posibles investigadores argentinos.

---

[17] Sobre la importante actuación de Aguilar en la medición de un arco de meridiano sobre territorio argentino, ver (Ortiz, 2005).

A principios de la década de 1930 la polémica sobre el Observatorio giraba alrededor de las actuaciones de su director, el astrónomo estadounidense Dr. Charles Dillon Perrine (1867-1951), a quién se le hacían diversas críticas y se intentaba desplazarlo de su cargo. Una de esas críticas se basaba en que, aunque Perrine había montado un excelente taller de óptica en Córdoba, no había logrado finalizar aún con el pulido de un espejo reflector de grandes dimensiones. Este era esencial para montar un instrumento que, quizás, podía llegar a convertirse en 'el gran telescopio argentino'.

Loyarte apreciaba a Aguilar y una vez incorporado a la Academia de Ciencias de Buenos Aires propuso que aquél fuera invitado a unirse a esa institución. Más tarde, a principios de la presidencia de Agustín P. Justo, cuando el Parlamento fue restituido Loyarte, en su carácter de Diputado Nacional, presentó una moción de crítica de las actuaciones de Perrine en el Observatorio de Córdoba (*Diario de Sesiones*, 1932). En su moción se solidarizó con las críticas formuladas anteriormente por Aguilar y por su colega Norberto Cobos en un conocido informe (Cobos y Aguilar, 1927; Minniti y Paolantonio, 2013**)**. Uno de los motivos de crítica fue la aparente falta de resultados positivos de las expediciones que ese observatorio había enviado al Brasil en 1912 (Barboza, 2010) y luego a Crimea (en 1914) y a Venezuela (en 1916). El objetivo de esas expediciones era observar eclipses totales de Sol y fenómenos ligados con ese evento; en particular, esos eclipses de Sol habrían podido verificar, o no, la validez de las ideas de Einstein relacionadas con la relatividad y su predicción de la deflexión de la luz de estrellas lejanas en su paso por cercanías del limbo solar. En todos los casos los objetivos de la expedición argentina, lo mismo que los de los otros países que enviaron misiones de observación, fueron frustrados por condiciones meteorológicas adversas (Perrine, 1923).

Aunque el Poder Ejecutivo formalmente rechazó la moción de crítica a Perrine, era sólo una cuestión de tiempo antes de que ese astrónomo fuera desplazado de la dirección del Observatorio Nacional (Ortiz, 2019)**.** Una vez que este hecho se concretó, la dirección quedó interinamente en manos de su principal crítico, Aguilar, que sin duda era entonces la figura local de mayor relieve en esa área. Sin embargo, tanto el nuevo director, como más tarde su sucesor, aceptaron la real dimensión del problema que aquejaba a Perrine en la configuración y pulido final del espejo en la Argentina: prudentemente lo enviaron a un establecimiento óptico muy especializado de los Estados Unidos. Una vez realizada esa tarea la lente fue devuelta desde Pittsburgh al Observatorio de Córdoba donde quedó instalada (Bernaola, 2001, pp. 247-260).

Sin embargo, puede afirmarse que, con la colaboración parlamentaria de Loyarte, la tendencia que favorecía el paso de la dirección de grandes institutos científicos a manos de especialistas locales fue robustecida. En esos años el proceso de transferencia de responsabilidades no se limitó solamente a los establecimientos de investigación científica, observatorios, museos o institutos: una historia paralela tuvo lugar dentro de organismos técnicos del Estado, incluyendo a los de las fuerzas armadas (Ortiz, 2015).

## 2.4 La actividad pública de Loyarte en la década de 1940

A pesar de la intensa actividad política que desarrolló hasta sus últimos años, Loyarte no abandonó su contacto con instituciones de promoción de la ciencia, en particular con la *Sociedad Científica Argentina*. En 1928 la SCA lo incorporó a su *Comisión de Anales* como editor responsable del área de la física; en esos mismos años Dassen era editor responsable del área de la matemática, y Damianovich de la de química. Diez años más tarde esa misma sociedad eligió a Loyarte vicepresidente primero por el período 1938-39; luego lo incorporó a su *Consejo Científico* como representante para el área de la física.

En la década siguiente, en 1942, Loyarte fue elegido nuevamente Diputado Nacional por el Partido Conservador. Un año más tarde, en junio de 1943, se produjo una segunda sublevación militar y el Congreso Nacional fue nuevamente disuelto. Aunque afectado por esa medida, Loyarte fue invitado a colaborar con las autoridades surgidas de ese segundo golpe militar y lo hizo en el área de la educación pública.

En 1884, cuando la emigración masiva y regular desde Europa hacia la Argentina estaba consolidándose, el Congreso Nacional sancionó la Ley 1420, conocida como *Ley de Educación Común*. Esa ley separaba de la educación oficial toda diferencia por credos religiosos, asegurando una enseñanza laica, gratuita y obligatoria en todo el territorio nacional. Consciente de la importancia de esa ley el entonces presidente de la Nación, Julio A. Roca (1843-1914), invitó a dos expresidentes, Domingo F. Sarmiento (1811-1888) y Nicolás Avellaneda (1837-1885), a cooperar en la elaboración del formidable proyecto educativo que se estaba formulando. Poco más tarde se creó el *Consejo Nacional de Educación* cuyo cometido era centralizar y uniformar el nuevo sistema educativo y coordinar el funcionamiento de las diferentes escuelas del Estado. Además, se crearon varias instituciones culturales colaterales, dependientes todas del Consejo. Una de ellas fue la *Biblioteca Nacional de Maestros*, en Buenos Aires, que es hoy una de las bibliotecas públicas mejor dotadas de Argentina. En 1881 Sarmiento aconsejó, para orientación de los maestros, la publicación de una revista especializada en problemas de la educación. Esa revista quedó bajo su dirección y se llamó *El Monitor de la Educación Común*.

Las autoridades surgidas del golpe de 1943, que no compartían aquellas ideas, designaron ministro de Justicia e Instrucción Pública al destacado escritor Gustavo Martínez Zubiría (1883-1962).[18] El nuevo ministro designó Interventor[19] del Consejo Nacional de Educación al contralmirante Pedro Gully pero, en muy poco tiempo, éste se alejó por razones de principios; para reemplazarlo Martínez Zubiría invitó a Loyarte.

Al instalarlo en su nuevo cargo, el 23 de octubre de 1943, el ministro destacó que Loyarte era conocido como un hombre 'distinguido por su cultura y su patriotismo' (*El Monitor*, 1943, p. 105); al mismo tiempo hizo explícitos los objetivos a los que aspiraba, en los campos de la educación y de la cultura, el movimiento militar que recientemente había tomado el poder en la Argentina. Martínez Zubiría explicó que: 'La escuela no puede, ni debe, ser solamente el aula de las primeras letras, con sus problemas elementales; la escuela debe ser la prolongación del hogar, y el hogar el complemento de la escuela, con su cotidiana contribución de amor y de ética y con la fervorosa invocación a Dios, fuente de toda virtud y perfección. Porque en la escuela debe reinar un ambiente Cristiano'. Más adelante precisó que 'todo ello ha de llevarse a cabo bajo el signo de la más pura argentinidad, para que la escuela sea como la soñaron los constructores del 53, el crisol donde los hijos de todos los hombres del mundo, que quieran habitar nuestro suelo, aprendan a amar y a morir por un solo Dios y por una sola bandera' (*El Monitor*, 1943, p. 106). Los principios de la enseñanza universal, laica y gratuita, introducidos por la *Ley de Educación Común* de 1884, y respetados desde entonces, debían ser reconvertidos y adaptados a las nuevas circunstancias políticas que Martínez Zubiría había descripto con tanta claridad. Esa fue la difícil tarea encargada a Loyarte, y que Gully no parece haber estado dispuesto a realizar.

---

[18] En sus obras literarias, que en su época tuvieron un impacto considerable, Martínez Zubiría usó el pseudónimo de 'Hugo Wast'.

[19] Funcionario designado directamente por el gobierno, en este caso por el gobierno surgido de ese golpe de Estado militar.

En el mismo acto Loyarte fue más moderado en sus palabras y, primeramente, dirigió un saludo respetuoso 'a los maestros' reconociendo la importancia central de su labor. Agregó que 'anticipándome a los sucesos, abrigo la certidumbre de que los maestros, ante la evidencia de la inspiración Cristiana que ha de guiar todos mis actos' se esforzarían y superarían en sus tareas dentro de un ambiente de comprensión (*El Monitor*, 1943, p. 107). Poco después designó a dos científicos de la UNLP como secretarios de su equipo: el matemático Agustín Durañona y Vedia y el bioquímico Carlos A. Sagastume.

Sin embargo, Loyarte no sería quién lograría llevar adelante esas políticas; se encontraba ya seriamente enfermo y hacia fines del año su salud comenzó a resentirse, lo que lo obligó a distanciarse de la Universidad[20] y dejar la dirección del Instituto de Física, posición que había conservado mientras fue presidente de la UNLP y luego Diputado Nacional. Poco más tarde esa misma dolencia le impidió participar en las celebraciones organizadas por él y por sus colaboradores para celebrar las Navidades de 1943. Pocos meses más tarde se vio obligado a renunciar también al cargo de Interventor del Consejo Nacional de Educación. Falleció el 30 de mayo de 1944, poco después de cumplir 56 años de edad.

**Consideraciones finales**

En este trabajo sobre el físico argentino Ramón G. Loyarte y el contexto científico y social en el que vivió, nos hemos enfocado principalmente en dos aspectos. El primero fue analizar, lo más objetivamente posible, el origen y el desarrollo de una célebre *polémica* que surgió cuando su investigación relacionada con la introducción de un nuevo número cuántico fue seriamente cuestionada. Como señalamos, fue un joven miembro del Instituto de Física bajo su dirección quien demostró la invalidez de los resultados de Loyarte. El segundo aspecto que desarrollamos en este trabajo fue la prolífica actividad política y partidaria que demandó la atención, la responsabilidad y el tiempo de Loyarte, especialmente en las últimas décadas de su vida.

Sabemos que Loyarte se destacó por su esfuerzo por impulsar trabajos originales en áreas de la física experimental, ya sea él solo o en colaboración con alumnos y miembros del Instituto de Física. Muchos resultados de esas investigaciones terminaron publicados en revistas nacionales y también en revistas extranjeras de entre las más leídas por sus pares. Algunos, particularmente aquellos que se refieren a mediciones experimentales precisas, continuaron siendo citados hasta fechas relativamente recientes.

Sin embargo, como vimos en el presente trabajo, un tópico central en las investigaciones de Loyarte, que involucraba la posible definición de un nuevo número cuántico y que hubiera tenido resonancia internacional, resultó ser incorrecto. El análisis de la validez real de esos trabajos generó una agria polémica local que comenzó en la revista científica de la UNLP y que luego se trasladó a *Physikalische Zeitschrift*, una conocida revista internacional de física.

Para presentar un análisis contextualizado y contemporáneo de los descubrimientos de Loyarte, y también de los aportes de los críticos de sus trabajos, hemos utilizado una herramienta novedosa en este género de estudios: los comentarios publicados en revistas internacionales de reseña científica. Esas revistas trataban de comunicar, tan objetivamente como fuera posible, el contenido científico y las ideas del trabajo reseñado, orientando a los lectores en sus búsquedas de bibliografía contemporánea sobre un tema dado. Los redactores de esas recensiones eran investigadores respetados dentro de su área específica; su función era comunicar las novedades del aporte, sin necesariamente expresar aprobación o rechazo. Sin embargo, muy excepcionalmente, algunos reseñadores no pudieron resistir al asombro que les causaba la comunicación de algún resultado: lamentablemente, esto ocurrió con algunos de los trabajos considerados en nuestro estudio.

---

[20] Loyarte, Legajos, Archivo General, Universidad Nacional de La Plata.

Éste y otros conflictos contemporáneos, en otras ramas de la ciencia en la Argentina de esos años que obviamente quedan fuera de este trabajo, y las respuestas que generaron localmente, sugieren un período de su desarrollo en el que el clima local parecía esperar de sus investigadores científicos solamente descubrimientos espectaculares. Consecuentemente, la realización de su verdadera dimensión dio lugar a frustraciones y aun a conflictos serios.

Como señalamos más atrás, Loedel Palumbo, el autor del agudo análisis que permitió demostrar que las hipótesis de Loyarte eran inválidas, era entonces uno de los investigadores jóvenes más destacados que había logrado formar el Instituto de Física de La Plata bajo la dirección de Gans, y luego de Loyarte. Lamentablemente, luego del episodio al que hacemos referencia, el futuro de este investigador dentro de su institución no fue auspicioso. Como ocurría en otras áreas de la vida ciudadana en esos años, quedó demostrado que el costo de la disidencia era alto.

Por otra parte, Loyarte, vinculado desde joven con movimientos políticos conservadores fue, también, el primer físico argentino y uno de los muy pocos científicos argentinos elegido, en varios períodos, Diputado ante el Congreso Nacional. Desde esas posiciones contribuyó a la sanción de la legislación que defendía los derechos de autor y de aquella que atribuía derechos de salud a los niños. Compartió también la sanción de una legislación represiva y se opuso a la admisión de refugiados políticos en un momento histórico muy crítico para Europa y para el mundo.

En los últimos años de su vida, en un período en el que la Argentina vivía nuevamente bajo un régimen militar, Loyarte se acercó a los nuevos gobernantes, quienes lo designaron Interventor del Consejo Nacional de Educación y le encargaron la difícil tarea de adecuar la educación básica, universal, gratuita y laica desde 1884, a nuevos principios que respondían a una visión menos tolerante de la composición de la sociedad argentina. Loyarte falleció sin haber podido cumplir con esa tarea.

**Agradecimientos**